# Low-Cost Optoelectronic Sensor for Early Screening of Citrus Greening in Leaves

[1]Ramji Gupta, [2]Ashis Kumar Das, [3]Sushmita Mena, [4]Saurav Bharadwaj

[1,4]Parul Institute of Engineering and Technology, Parul University, Vadodara, Gujarat, India

[2]ICAR-Central Citrus Research Institute, Nagpur, Maharashtra, India

[3]Parul Institute of Applied Sciences, Parul University, Vadodara, Gujarat, India

[4]Corresponding author: Saurav Bharadwaj, saurav.bharadwaj33162@paruluniversity.ac.in

**Abstract:** Citrus greening, or Huanglongbing (HLB), is a serious disease affecting citrus crops, with no known cure. Early detection is essential, but current methods are often expensive. To address this, a low-cost, portable sensor was developed to distinguish between HLB-infected and healthy citrus leaves using a LED-based optical sensing circuit. The device uses white and infrared (IR) LEDs to illuminate the adaxial leaf surface and measures change in reflectance intensities caused by differences in biochemical compositions between healthy and HLB-infected leaves. These changes, analyzed across four spectral bands (blue, green, red, and IR), were processed using machine learning models, including Random Forest. Experimental results indicated that the IR band was the most effective, with the Random Forest model achieving an accuracy of 89.58% and precision of 93.75%. Similarly, the green band also achieved an accuracy of 85.42% and precision of 90.62%. These results suggest that this LED-based optical system could be a hand-held screening tool for early detection of HLB, providing small-scale farmers with a cost-effective solution.

**1 Introduction**

Citrus greening, also known as Huanglongbing (HLB), is a devastating bacterial disease that poses a significant threat to citrus trees worldwide. It is caused by a Gram-negative bacterium from the genus *Candidatus Liberibacter asiaticus*, which belongs to the Rhizobiaceae family. This bacterium targets the phloem tissues of citrus trees, disrupting nutrient transport and inducing severe physiological changes [1], [2]. The disease is spreaded by the Asian citrus psyllid (*Diaphorina citri*), which serves as its primary vector. Symptoms of citrus greening include chlorosis, where leaves turn yellow while the veins remain dark green. Infected trees develop smaller, narrower leaves and produce misshapen fruits that are bitter with reduced juice content [3]. As the disease progresses, tree growth is stunted, branches die back, and the tree may ultimately perish. Symptoms typically appear months to over a year after infection, allowing the disease to spread across the farm through vectors before being detected [4]. Disease is highly destructive, causing 30–100% yield losses, rendering trees unproductive within 2–5 years of symptom onset, and shortening their lifespan to just 7–10 years [5]. Currently, there is no cure for citrus greening. Early detection and robust management strategies are essential to mitigate its spread [6]. Prompt removal of infected trees helps prevent further orchard contamination. Controlling the psyllid vector through insecticides and biological measures is critical, while optimal nutrition and irrigation practices can improve the resilience of citrus trees against this destructive disease [7].

Visual inspection serves as an initial method for detecting citrus diseases. Expert scouts conduct meticulous examinations of orchard trees, focusing on signs and symptoms observed on fruits and leaves. These trained professionals identify infected trees based on visible symptoms [8], [9]. This approach relies on a thorough assessment of the physical manifestations of disease. However, visual inspection is limited by its inability to detect diseases in asymptomatic trees [6]. The high costs associated with employing expert scouts and the risk of missing latent infections underscore the need for more reliable disease-screening tools. This is particularly critical on large-scale farms, where examining each plant individually is both time-consuming and labour-intensive [7].

Molecular techniques are crucial for the detection and diagnosis of citrus greening disease. The polymerase chain reaction (PCR) is used to amplify specific DNA sequences of the pathogen, employing primers that target conserved regions of the bacterial genome, thereby enabling qualitative detection. Real-time PCR goes a step further by quantifying the bacterial DNA in citrus leaves, providing both detection and quantification of the pathogen [1]. The LAMP technique offers a highly specific detection method under constant temperature conditions, eliminating the need for thermal cycling [10], [11]. Next-generation sequencing technologies allow comprehensive analysis of microbial communities by directly sequencing bacterial genomes from infected samples [12]. Additionally, immunological assays such as ELISA are employed to detect bacterial antigens in tissues, serving as a complementary approach to molecular techniques for disease diagnosis [13]. Despite their effectiveness, these advanced methods require meticulous sample preparation, significant time, and specialized expertise, making them impractical for routine use by the small-scale farmer [7], [14].

Researchers utilized portable visible-near-infrared (vis-NIR) spectroscopy as a rapid and non-destructive optical screening tool for real-time monitoring of disease stress in citrus leaves [7], [15]. Field-based near-infrared (NIR) spectroscopy, combined with various machine learning algorithms for spectral analysis, enabled the detection of citrus greening disease. Important wavelengths, such as 570 nm and 670 nm in the visible spectrum and 870 nm and 970 nm in the NIR spectrum, have been demonstrated to be crucial markers of disease stress in the citrus canopy [16], [17], [18], [19]. Several machine learning models were employed to classify spectral data. Among these, the support vector machine (SVM) model demonstrated the highest classification accuracy, achieving 97% in distinguishing infected spectra from healthy ones [8]. Selected spectral bands at 537 nm, 662 nm, and 713 nm in visible and 813 nm, 1120 nm, and 1472 nm in NIR ranges were employed; this resulted in an average classification accuracy of 85% [15], [20].

Modern spectroscopes require significant initial investment and pose considerable challenges for widespread adoption among citrus farmers [7]. Field conditions, such as temperature fluctuations, humidity variations, and light interference, can adversely impact the accuracy of reflectance measurements [17]. Developing and maintaining precise calibration models for spectroscopy is both time-intensive and complex, necessitating frequent updates with new field data to account for environmental variability [18]. Moreover, the analysis and interpretation of spectral data demand specialized knowledge and technical expertise [13], [15]. Field-based spectrometers often require regular recalibration and maintenance to ensure consistent performance under challenging field conditions. Currently, no commercially available portable tools can accurately predict citrus greening disease in citrus leaves for farmers [4].

In this study, a cost-efficient, non-invasive, and non-destructive LED-based sensor technique was developed to differentiate between infected and healthy citrus leaves. The sensor system employs white and IR light emitting diodes (LEDs) to illuminate the adaxial leaf surface, facilitating the capture of reflectance intensity variations that result from differential absorption characteristics of the leaf biochemicals. The core objective of this research is to assess whether these miniaturized sensor circuits can effectively identify reflectance changes corresponding to the absorption peaks of specific biochemicals associated with disease status. By leveraging the unique signatures obtained from the reflectance data, the sensor aims to accurately classify the leaves as either infected or healthy. This approach is intended to provide a practical, portable tool for field-based diagnostics, enabling efficient disease-stress assessment in citrus trees without the need for extensive laboratory analysis. The anticipated outcome is a rapid and reliable method for monitoring plant health, thereby supporting timely intervention and management strategies in citrus cultivation.

## 2 Materials and Methods

### 2.1 Leaf Samples

Seven healthy and seventeen HLB-infected leaves, exhibiting blotchy mottle and green island symptoms (typical of HLB), were collected from a sweet orange (*Citrus sinensis* cultivar Mosambi) tree in an orchard located in the Nagpur district of Maharashtra, India, as shown in Figures 1. The branches from which the leaves were taken were confirmed to be HLB-positive through laboratory-based quantitative PCR (qPCR) tests, as described previously [1], [21]. Each leaf was gently cleaned with a soft brush to remove any adhering dust particles. After cleaning, the leaves were sealed in ziplock bags to preserve their integrity and prevent moisture loss. They were subsequently stored in a refrigerator at a controlled temperature between 0°C and 5°C, with relative humidity maintained between 30% and 50%. These storage conditions were critical for preserving the biochemical properties of the leaves and preventing degradation during subsequent examinations. Ten reflectance measurements were randomly taken from different adaxial spots on each leaf.

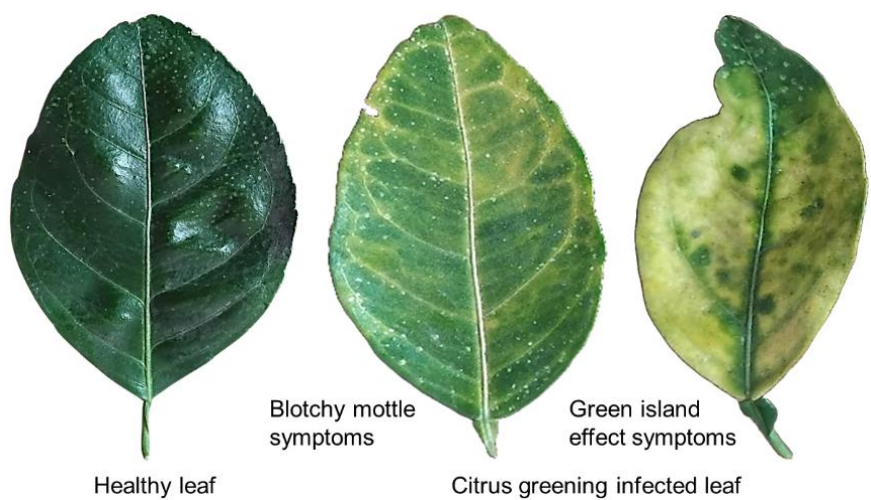

Figure 1: Sweet orange leaves collected from Nagpur district of Maharashtra.

**2.2 Sensor Circuit Design**

Reflectance intensities from the adaxial surface of the leaves were recorded due to its higher pigment concentration compared to the abaxial surface. White and IR LEDs were used to excite the leaf samples, and the reflected intensities were acquired and analysed using multiple sensors interfaced with an Arduino Nano (ATmega328) microcontroller. The microcontroller has 32 KB of flash memory, 2 KB of SRAM, and 1 KB of EEPROM. It runs at a clock speed of 16 MHz. Colour and IR sensors are the two separate sensor modules that make up the circuit.

The color sensor (TCS34725) is equipped with two integrated white LEDs that emit light across the 380–700 nm wavelength range, with an intensity of 600–800 lux as measured using an LX-103 digital light meter. It measures the reflectance intensities for blue (465 nm), green (525 nm), and red (615 nm) using photodiodes and digitizes the data with an ADC, producing a 16-bit cumulative intensity value that is stored in a register, as shown in Figures 2(a) and 2(c). Communication with the sensor is achieved through the I2C protocol, utilizing bidirectional open-drain lines SDA and SCL, which are pulled high via pull-up resistors. The sensor interfaces with the microcontroller through the SDA and SCL pins, which are connected to the A4 and A5 pins, respectively. Data is transmitted on the SDA line and synchronized with clock pulses on the SCL line, with each data bit transmitted during the low phase of the clock cycle.

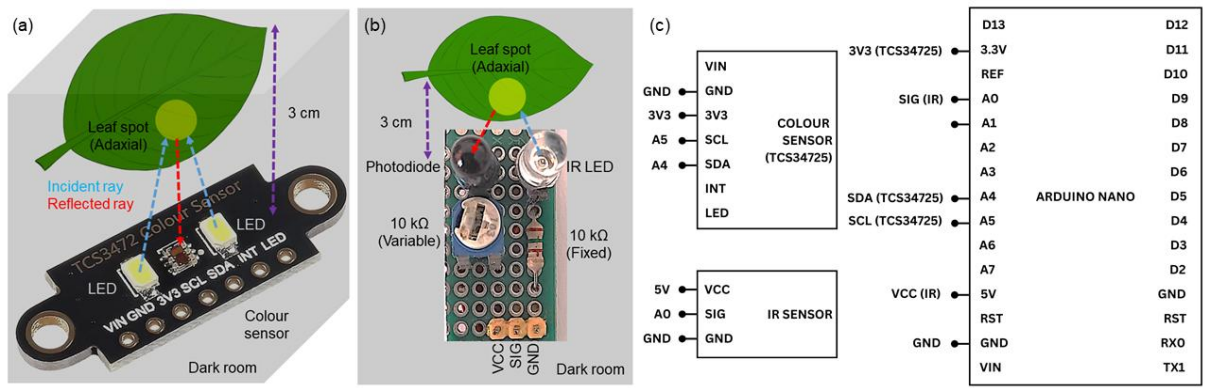

Figure 2: (a) Colour and (b) IR reflectance measurement of leaves. (c) Schematic diagram of the sensor circuit.

The custom-designed IR sensor module, with compact dimensions of 30 mm × 15 mm, incorporates a 5 mm IR emitter and a 5 mm IR photodiode configured for optimal performance within the 700–1000 nm spectral range, peaking at 940 nm, as illustrated in Figures 2(b) and 2(c). The photodiode is connected to the A0 analog pin of the microcontroller, enabling the conversion of varying reflected IR light intensities into corresponding voltage signals. Variations in reflected light intensity lead to changes in the photodiode resistance, which in turn modify the output voltage. The output voltage is directly proportional to the intensity of the reflected IR light.

**2.3 Machine Learning Techniques**

In Figure 3(a), the raw sensor output was first subjected to min–max normalization, linearly scaling every feature to the closed interval [0, 1]. After normalization, three supervised learning algorithms—k-Nearest Neighbors (k-NN), decision tree, and random forest—were trained to discriminate between healthy and infected leaves. The full dataset was stratified into independent training and test partitions in an 80 : 20 ratio to enable unbiased performance estimation.

For the k-NN model, k was fixed at 3. Each unknown sample was assigned the class most common among its three nearest neighbors in the Euclidean feature space, although alternative distance metrics remain configurable for future sensitivity analyses. The method presupposes that members of the same class form compact clusters; therefore its efficacy is maximal when the manifold of healthy and diseased spectra is well separated.

The decision-tree classifier performed binary partitioning by recursively selecting the feature and threshold that maximized information gain, evaluated via Gini impurity or entropy. Tree growth halted upon reaching a predefined maximum depth or when further splits failed to increase class purity beyond a user-specified threshold. Post-training, the pruned tree was validated on the hold-out set to confirm that the learned decision rules generalize and do not overfit spurious patterns in the training data.

A random-forest ensemble of 500 bootstrap-sampled trees was constructed. During node splitting, $\sqrt{p}$ randomly chosen features (where p is the total number of predictors) were considered, injecting de-correlation among trees and reducing variance relative to a single tree. Leaf-level predictions were aggregated by majority vote to yield the final class label. Evaluation on the test partition quantified the resilience and its capacity to capture non-linear interactions that individual trees or k-NN might miss.

## 3 Results and Discussion

### 3.1 Visual Examination

Blotchy mottle is a prominent symptom of HLB, characterized by irregular patterns of chlorosis and green pigmentation on infected citrus leaves. These patterns are asymmetrical and randomly distributed across the leaf surface, unlike the symmetrical chlorosis observed in nutrient deficiencies that typically align along the leaf midrib [22]. The lack of uniformity in blotchy mottle serves as a diagnostic feature distinguishing it from other abiotic stresses. The underlying mechanism for blotchy mottle involves the colonization of phloem sieve tubes by the HLB-associated bacteria [23]. These pathogens disrupt the vascular system by proliferating and blocking the phloem, thereby impeding the transport of photosynthates and essential nutrients. The resulting physiological imbalances manifest as localized nutrient deficiencies or toxic accumulations, creating contrasting chlorotic and green regions on the leaves. This disruption highlights the bacterial impact on the plant ability to maintain uniform nutrient distribution and chlorophyll content, further exacerbating symptoms of asymmetrical chlorosis [24]. The severity of blotchy mottle varies depending on factors such as leaf age, citrus variety, and the stage of infection. Younger leaves or those from highly susceptible varieties often exhibit more pronounced mottling. As the infection progresses, the intensity of blotchy mottle correlates with increasing physiological stress and overall decline in plant health. Thus, blotchy mottle serves not only as a visual marker for HLB but also as an indicator of disease progression and severity [7], [25].

Green island symptom is another distinct feature of HLB-affected citrus leaves, characterized by patches of green, healthy-looking tissue amidst surrounding yellow or chlorotic areas. These green islands arise due to localized physiological and biochemical disruptions caused by the disease, resulting in the uneven degradation of chlorophyll across the leaf surface [26]. This phenomenon is attributed to the localized retention of photosynthetic activity in the green areas, driven by alterations in the metabolic pathways of plants. The pathogen-induced disruption of normal signaling mechanisms for chlorophyll breakdown is a key factor in green island formation. Specifically, the accumulation of plant hormones such as cytokinins in the affected areas inhibits chlorophyll degradation, preserving the green pigmentation in these regions [27]. Cytokinins are known to play a role in delaying senescence and maintaining chloroplast function, which likely explains the persistence of these green patches. Simultaneously, nutrient transport to other parts of the leaf is compromised due to the bacterial colonization of phloem, leading to chlorosis in the surrounding tissue. Green islands are diagnostic of HLB and illustrate the complex interplay between the pathogen and the host physiological processes [28]. The contrasting appearance of chlorotic and green regions underscores the systemic impact of HLB on the plant, reflecting the pathogen ability to manipulate host metabolism at localized levels. This symptom not only aids in identifying the disease but also provides insight into its underlying biochemical and hormonal disruptions, offering valuable information for understanding the pathogen-host interaction [7], [29].

Asymptomatic leaves are from citrus trees infected with the HLB-causing bacterium, but they do not exhibit the typical visual symptoms associated with the disease. These leaves appear similar to healthy leaves, making them challenging to identify based solely on appearance. The presence of the pathogen can be confirmed through molecular methods such as qPCR [25].

## 3.2 Classification

The classification performance of k-NN, Decision Tree, and Random Forest models was evaluated across four distinct spectral bands—blue, green, red, and IR—using standard metrics such as accuracy, precision, recall, and F1-score. For the k-NN model, the red band exhibited the highest classification capability with an accuracy of 0.8750, precision of 0.9355, the greatest recall value of 0.9062, and an F1-score of 0.8788. The IR band followed closely, achieving an accuracy of 0.8542, precision of 0.9062, recall of 0.8923, and an F1-score of 0.8788. The blue band demonstrated strong performance with an accuracy of 0.8542, precision of 0.9333, recall of 0.8889, and an F1-score of 0.8588, while the green band yielded an accuracy of 0.8125, precision of 0.8750, recall of 0.8615, and an F1-score of 0.8683 (Table 1).

The Decision Tree model showed its best performance in the IR band, attaining an accuracy of 0.8750, precision of 0.9355, recall of 0.9062, and an F1-score of 0.8788, highlighting its superior classification ability among the bands. The green band also yielded high performance with an accuracy of 0.8542 and a precision of 0.9333, while both recall and F1-score were measured at 0.8889 and 0.8485, respectively. The blue band exhibited balanced performance with an accuracy of 0.7917 and equal precision, recall, and F1-score values of 0.8485. The red band demonstrated relatively lower classification ability, showing an accuracy of 0.7708, precision of 0.8929, recall of 0.8197, and an F1-score of 0.7576.

In the case of the Random Forest model, the IR band again demonstrated the highest classification performance, with an accuracy of 0.8958, precision of 0.9375, recall of 0.9231, and an F1-score of 0.9091. The green band followed, yielding an accuracy of 0.8542, precision of 0.9062, recall of 0.8923, and an F1-score of 0.8788. The red band achieved an accuracy of 0.8125, precision of 0.8750, recall of 0.8615, and an F1-score of 0.8485, while the blue band maintained a balanced classification performance with accuracy, precision, recall, and F1-score all at 0.7917 and 0.8485, respectively. These Random Forest results were further analyzed graphically to provide deeper insights into model behavior and spectral sensitivity.

From a physiological perspective, the superior performance of the IR band may be attributed to its sensitivity to starch accumulation in citrus leaves. Although citrus leaves typically contain low levels of starch, abnormal accumulation can occur due to phloem blockages caused by bacterial infections, such as Huanglongbing (HLB). This buildup disrupts chloroplast function, leading to leaf yellowing and branch dieback, making elevated starch levels a strong indicator of HLB [30]. On the other hand, the green band reflects variations in chlorophyll content, which decline during disease progression. Chlorosis, resulting from nutrient deficiencies or hormonal imbalances, leads to decreased photosynthetic capacity, early leaf abscission, and reduced fruit yield and quality. Thus, the extent of chlorophyll loss in the green band serves as a valuable marker for assessing disease severity [31].

Table 1: Performance metrics of different algorithms across wavelengths.

| Algorithm | Wavelength | Training Metrics | | | | Testing Metrics | | | |
|---|---|---|---|---|---|---|---|---|---|
| | | Accuracy | Precision | F1 Score | Recall | Accuracy | Precision | F1 Score | Recall |
| K-Nearest Neighbor | Blue | 0.8229 | 0.8815 | 0.8686 | 0.8750 | 0.8542 | 0.9333 | 0.8485 | 0.8889 |
| | Green | 0.8854 | 0.9137 | 0.9270 | 0.9203 | 0.8125 | 0.8750 | 0.8485 | 0.8615 |
| | Red | 0.8854 | 0.9078 | 0.9343 | 0.9209 | 0.8750 | 0.9355 | 0.8788 | 0.9062 |
| | IR | 0.8438 | 0.8741 | 0.9124 | 0.8929 | 0.8542 | 0.9062 | 0.8788 | 0.8923 |
| Decision Tree | Blue | 0.8490 | 0.9219 | 0.8613 | 0.8906 | 0.7917 | 0.8485 | 0.8485 | 0.8485 |
| | Green | 0.9219 | 0.9621 | 0.9270 | 0.9442 | 0.8542 | 0.9333 | 0.8485 | 0.8889 |
| | Red | 0.8906 | 0.9394 | 0.9051 | 0.9219 | 0.7708 | 0.8929 | 0.7576 | 0.8197 |
| | IR | 0.8594 | 0.8929 | 0.9124 | 0.9025 | 0.8750 | 0.9355 | 0.8788 | 0.9062 |
| Random Forest | Blue | 0.8490 | 0.8750 | 0.9197 | 0.8968 | 0.7917 | 0.8485 | 0.8485 | 0.8485 |
| | Green | 0.9219 | 0.9420 | 0.9489 | 0.9455 | 0.8542 | 0.9062 | 0.8788 | 0.8923 |
| | Red | 0.8906 | 0.9143 | 0.9343 | 0.9242 | 0.8125 | 0.8750 | 0.8485 | 0.8615 |
| | IR | 0.8594 | 0.8767 | 0.9343 | 0.9046 | 0.8958 | 0.9375 | 0.9091 | 0.9231 |

**3.3 Graphical Analysis**

Training and testing curves are critical tools for evaluating the performance and generalization capability of Random Forest models. As illustrated in Figure 3(d), the variation in performance metrics as a function of training data volume reveals significant trends. Specifically, as the amount of training data increases, training accuracy tends to decrease slightly due to reduced overfitting, while testing accuracy improves, indicating enhanced generalizability. Upon training with the full dataset, both training and testing accuracies stabilize, marking the learning saturation point. These observations underscore the importance of sufficient training data in achieving reliable and accurate model predictions.

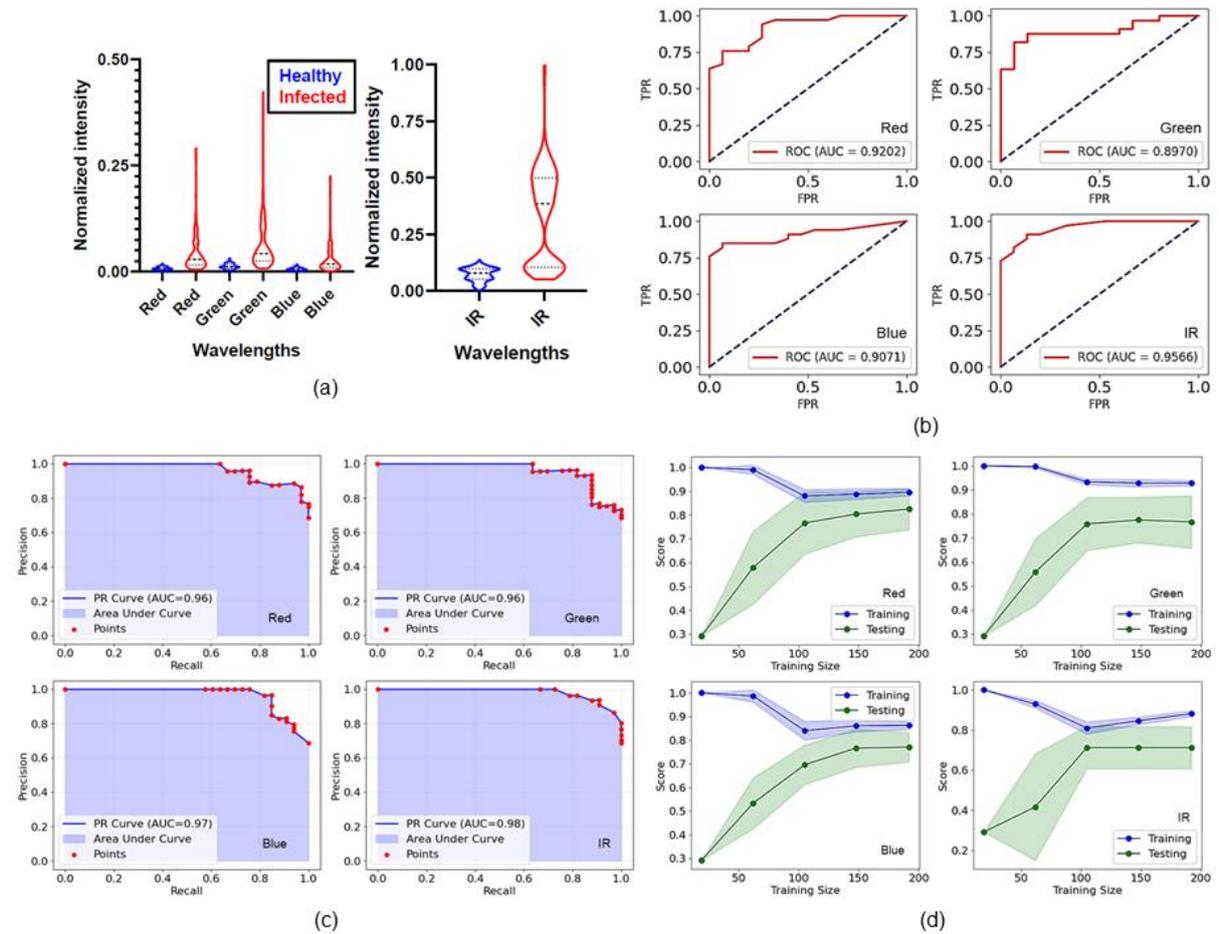

Figure 3: (a) Normalized reflectance intensities across varying wavelengths. (b) ROC curves with AUC values for Random Forest models, indicating classification performance. (c) PR curves with AUC values for the same models, emphasizing performance on imbalanced datasets. (d) Learning curves showing training and validation scores of Random Forest models as a function of training set size.

Random Forest models were evaluated using Receiver Operating Characteristic (ROC) curves along with the corresponding Area Under the Curve (AUC) values across multiple spectral bands. The ROC curve represents the diagnostic ability by plotting the True Positive Rate (TPR) against the False Positive Rate (FPR), thereby illustrating its discrimination power between healthy and HLB-infected samples. The AUC quantifies this performance, with higher values reflecting better distinction between classes. The AUC scores obtained for the red (0.9202), green (0.897), blue (0.9701), and infrared (0.9566) bands indicate strong classification performance across all spectral domains, as depicted in Figure 3(b).

Precision-Recall (PR) curves were utilized to evaluate the models in a binary classification context, offering insights into the trade-off between precision and recall at various threshold settings. The area under the PR curve (AUC-PR) serves as a key metric for assessing model effectiveness, especially under class imbalance. AUC-PR values nearing 1 suggest exceptional classification performance. In this study, the Random Forest models yielded AUC-PR values of 0.96 for Red, 0.96 for Green, 0.97 for Blue, and 0.98 for Infrared spectral bands. These results, visualized in Figure 3(c), affirm the robustness in distinguishing between HLB-infected and healthy cases with high precision and recall.

### 3.4 Comparison

Previous studies on the use of NIR spectroscopy for plant disease detection have explored a broad spectral range from 350 nm to 2500 nm. These investigations commonly employed high-end, research-grade instruments such as the SVC HR-1024 and FieldSpec 3. Although these devices demonstrated high classification accuracy using machine learning techniques, their integration into field-based commercial products has been limited. The reasons for this lack of adoption remain unclear. However, a probable factor could be the high cost of such spectroscopic equipment, which makes them economically unfeasible for small- and medium-scale farmers, especially for routine or rapid on-site diagnosis of HLB disease in citrus crops.

In response to this challenge, the proposed research presents a cost-effective, hand-held LED-based reflectance measurement device tailored for visible and near-infrared wavelengths. The prototype is designed to offer rapid screening capabilities at a significantly reduced cost, estimated to be under ₹5000, thereby enhancing accessibility for smallholder farmers. While the device may not match the precision of high-end spectrometers, its intended role as a preliminary screening tool justifies its performance. Suspected cases identified through this device can be subsequently verified using more precise diagnostic techniques such as PCR. The system was evaluated using the Random Forest classification algorithm and achieved an accuracy of 89% in distinguishing between healthy and HLB-infected citrus leaves within the infrared spectral band. This level of accuracy, while modest compared to lab-based instruments, demonstrates the feasibility of the device for field-level disease screening. As shown in Table 2, the proposed approach offers a practical balance between affordability and diagnostic utility, making it a viable tool for large-scale deployment in citrus orchards, particularly in resource-constrained agricultural settings.

Table 2: Comparison of historical and proposed research.

| Wavelengths Range (nm) | Technique | Device | Cost | Statistics | Accuracy |
|---|---|---|---|---|---|
| 350-2500 | Vis-NIR | SVC HR-1024 | High | QDA | 98% [8] |
| 537, 662, 713, 813, 1120, 1472 | Vis-NIR | SVC HR-1024 | High | QDA-SIMCA | 87% [20] |
| 570, 670, 870, 970 | Vis-NIR | Self-built | High | SVM | 97% [19] |
| 350-2500 | Vis-NIR | FieldSpec 3 | High | SVM | 97% [16] |
| 465, 525, 615, 940 | Vis-NIR | - | Low ($60) | Random Forest | 89% (Proposed) |

**4 Conclusion**

The research developed a compact, cost-efficient, and non-destructive sensor system that utilizes an optical sensing circuit to differentiate between healthy and infected citrus leaves. By employing white and IR LEDs, the system captures reflectance intensity variations based on the differential absorption characteristics of leaf biochemicals, enabling the detection of early symptoms of HLB disease. Three machine learning models were evaluated to assess the disease status of the leaves. Among these, the Random Forest model demonstrated the most effective performance across various spectral bands, with the IR band yielding the highest classification accuracy and precision. The IR band detects changes in starch levels in citrus leaves, and when coupled with the green band, which identifies chlorophyll variations, it offers valuable insights into symptoms of HLB, such as blotchy mottle and the green island symptom. These findings underscore the potential of optical sensing coupled with machine learning techniques for the precise, early detection of HLB, which could significantly aid in the management and monitoring of citrus crop health, ensuring better-targeted interventions and reducing the impact of this devastating disease on the citrus industry.

**Statements and Declarations:**

**Funding:** The project is funded by Parul University (RDC/IMSL/167).

**Conflict of interest:** The authors declare that they have no conflict of interest.